\documentclass[aps,pra,10pt,twocolumn,floatfix, footinbib,superscriptaddress]{revtex4-1}

\usepackage{graphicx, color, soul}
\usepackage{amsmath, amsfonts, amssymb, mathrsfs, dsfont, mathtools}
\usepackage[usenames,dvipsnames, pdftex]{xcolor}
\usepackage{ulem} 

\usepackage[mathscr]{eucal}

\usepackage[colorlinks, breaklinks, 
            linkcolor=OrangeRed,
            citecolor=RoyalBlue,
            urlcolor=RoyalBlue]{hyperref}
            
\usepackage{multirow}

\usepackage[all]{hypcap} 

\newcommand{\Le}{\left}
\newcommand{\Ri}{\right}

\newcommand{\f}{\frac}

\newcommand{\ra}{\rangle}
\newcommand{\la}{\langle}

\newcommand{\eq}[1]{\begin{align}#1\end{align}}

\newcommand{\msr}{\mathscr}

\newcommand{\thop}{t_\text{h}}
\newcommand{\trho}{\tilde{\varrho}}
\newcommand{\err}[1]{\delta^{\text{#1}}(t)}
\newcommand{\ua}{\underline{\alpha}} 
\newcommand{\ub}{\underline{\beta}} 
\newcommand{\ug}{\underline{\gamma}} 
\newcommand{\oI}{\bar{I}}

\begin{document}


\title{Dephasing in strongly disordered interacting quantum wires}

\author{Sourav Nandy}
\affiliation{Department of Physics, Indian Institute of Technology Bombay, Mumbai 400076, India}
\author{Ferdinand Evers}
\affiliation{Institute of Theoretical Physics, University of Regensburg, D-93040 Germany}
\author{Soumya Bera}
\affiliation{Department of Physics, Indian Institute of Technology Bombay, Mumbai 400076, India}

\date{\today}

\begin{abstract}
Many-body localization is a fascinating theoretical concept describing the intricate interplay of quantum interference, i.e. localization, with many-body interaction induced dephasing. Numerous computational tests and also several experiments have been put forward to support the basic concept. Typically, averages of time-dependent global observables have been considered, such as the charge imbalance. 
We here investigate within the disordered spin-less Hubbard ($t-V$) model how dephasing manifests in time dependent variances of observables. 
We find that after quenching a N\'eel state the local charge density exhibits strong temporal fluctuations with a damping that is sensitive to disorder $W$: variances decay in a power law manner, $t^{-\zeta}$, with an exponent $\zeta(W)$ strongly varying with $W$. A heuristic argument suggests the form, $\zeta\approx\alpha(W)\xi_\text{sp}$, where $\xi_\text{sp}(W)$ denotes the noninteracting localization length and $\alpha(W)$ characterizes the multifractal structure of the dynamically active volume fraction of the many-body Hilbert space.
In order to elucidate correlations underlying the damping mechanism, exact computations are compared with results from the time-dependent Hartree-Fock approximation. Implications for experimentally relevant observables, such as the imbalance, will be discussed. 
\end{abstract}

\maketitle

\section{Introduction}
Understanding the effect of interactions in low dimensional Anderson localized system has gained a lot of momentum in last few decades~\cite{Gornyi2005, Basko2006,Oganesyan2007, Nandkishore2015, Imbrie2016, AnnMBLReview2017, Alet2018, AbaninBloch-Review-2018}. 
Largely based on numerical evidence obtained, e.g., in the random-field Heisenberg (or $t-V$) model it is believed that under generic conditions in one dimension, even at finite temperature, a many-body localized (MBL) phase is stable being resilient against interaction induced dephasing effects~\cite{Oganesyan2007, Znidaric2008, Pal2010, Bardarson2012, Potter2015, BarLevPRL2015, Luitz2015, Bera2015, ModakPRL15, Znidaric2016, BeraArul16, Singh2016, MierzejewskiPRB16, NagPRB17, DeTomasi2017, Khemani2017, Bera2017, Loic2018, Lenarcic2018,  SierantPRB19,DetomasiPRB19,SierentPRL20}. 
Recently, signatures of MBL have also been reported in several experimental studies~\cite{Schreiber2015, Smith2016, Choi2016, Luschen2017, BordiaPRX2017, RoushanScience17, RispoliExp18, KohlertPRL2019, WeiNuclearSpin18, LukinScience19}.

For the $t-V$-model it is perhaps too early to declare consensus about the existence of an MBL-phase proper~\cite{Bera2017, PandaMBL19, SirkerPRL20},
i.e. an emergent integrable phase with local integrals of motions~\cite{Serbyn2013, Huse2014, Ros2015, Rademaker2016, Imbrie2016, Imbrie2016-2, OBrien2016}.
The computational challenge to overcome in the strong-disorder regime is the (expected) dynamical slowing down together with the (unexpected) strong effect of finite sample sizes \cite{Bera2017, Weiner19, ChandaPRB20}.
By now there is an overwhelming evidence that a large parameter regime exists exhibiting a very slow relaxation of conserved quantities~\cite{Luitz2016, Bera2017, Doggen2018, ChandaPRB20}. 
However, the detailed nature of dynamical phenomena in large-disorder (finite-energy-density) phases are still partially unexplored and mostly not understood. The overall situation is well illustrated by the fact that the critical disorder strength, $W_c$, for the transition into the MBL-phase is not accurately known. Current estimates for the Heisenberg model range between $3.8$ and $5.5$  - with computationally larger studies tending towards higher values~\cite{Devakul2015, Bera2017, Doggen2018, Lenarcic2018, PandaMBL19, ChandaPRB20, SierantLargeWc20}. 

The large spread in the estimate for $W_\text{c}$ may partially be explained by a recent conjecture: it is proposed that the ergodic phase with power-law dynamics for the width  of the diffusion propagator, $\Delta x(t)\sim t^\beta$,  is separated from the MBL-phase by an intermediate phase with an unbound growth of $\Delta x(t)$ slower than any power~\cite{Weiner19}. 
The intermediate phase  is situated within a window of disorder values $W_{\text{c}_1}\lesssim W \lesssim W_{\text{c}_2}$~\cite{Weiner19, Khemani2017}; the earlier work~\cite{Luitz2015} would be consistent with $W_{\text{c}_1}\approx 4$, while the more elaborate later estimates would hint at $W_{\text{c}_2}\gtrsim 5$~\cite{Devakul2015, Doggen2018, SierantLargeWc20}.

Here, we continue our numerical investigation of the $t-V$-model. From earlier studies we borrow the observation that at disorder values  $W\gtrsim 5$ the true asymptotic regime of charge dynamics is very hard to reach and may, in fact,  be situated at observation times and system sizes outside the window of ``numerically exact'' computations. We are thus motivated to search for signatures of MBL-associated physics that manifest already at shorter times and smaller system sizes. A promising sensor we here explore is the damping of time-dependent fluctuations; it may be analyzed by evaluating  quenches in terms of  ensemble-averaged variances of observables, such as the local density or the imbalance taken in a finite-size sample. Remarkably, within our window of observation times and at moderate to large disorder, temporal fluctuations exhibit a non-exponential, close to power-law decay $t^{-\zeta(W)}$, 
so that there is no simple notion of a single decay rate. Moreover, sample-to-sample fluctuations are large so that average and typical fluctuations decay with exponents differing by 35\% at moderate disorder. 
Our computations confirm that damping is indeed very sensitive to the disorder regime; it is large in the moderate-disorder / thermal phase ($\zeta(W)\approx 1$ at $W= 1.5$), 
while we find it to (nearly) vanish at stronger disorder 
indicating the expected lack of ergodicity.
Our predictions can be readily tested in contemporary experimental setups as they have been used before in the field. 
\cite{Schreiber2015, Choi2016, Luschen2017, BordiaPRX2017}.

\begin{figure*}[thb]
    \centering
    {\includegraphics[width=1.0\textwidth]{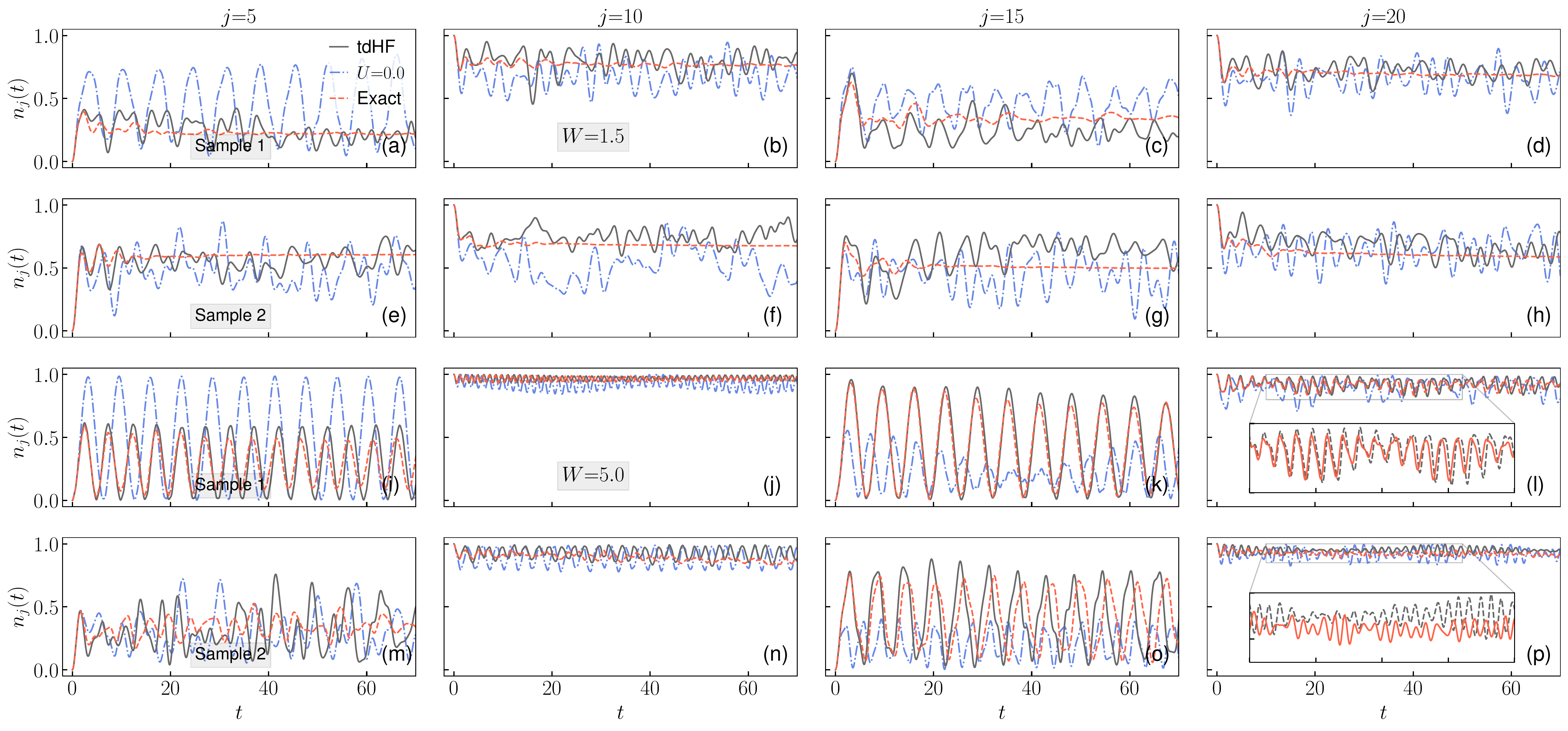}}
    \caption{Local charge density, $n_{j}(t)$, as a function of time at different sites $j$ for two disorder strengths, $W
    {=}1.5$ (rows one and two) and $5.0$ (rows three and four), at interaction strength $U
    {=}1.0$ for $L{=}24$ along with the non-interacting density. Two typical disorder configurations per $W$-value highlight the typical behavior, e.g., for differences between tdHF~(solid line) and exact~(dashed line) calculation. 
    The insets (l,p) highlights the deviation between exact and tdHF traces at intermediate time.
    \label{F1}}
\end{figure*}

In order to elucidate the physical origin of the damping mechanism, we compare results from exact traces with the 
time-dependent Hartree-Fock (tdHF) approximation\cite{ringschuck} using it as a diagnostic tool. The tdHF has been adopted previously in dynamical studies of MBL phases\cite{KnapHFPRB18}. A refined variant of tdHF, the second order Born approximation (SCBA), has also been  employed~\cite{BarLev14,Rajdeep19,BarlevEPL2016}.  
Our results indicate that tdHF-traces have a tendency towards equilibration even at large disorder implying that MBL-physics is not appropriately captured. This finding is  at variance with earlier reports~\cite{KnapHFPRB18}.

\section{Model and Methods}
We consider the paradigmatic $t{-}V$ model that describes a ring of spinless fermions with 
Hamiltonian
%
\begin{eqnarray}
\label {eq:H}
H &=& H^{(0)} + U \sum_{i=1}^{L-1}(n_i-1/2)(n_{i+1}-1/2), \\
H^{(0)} &\coloneqq& -\frac{\thop}{2}\sum_{i=1}^{L-1}c_{i}^{\dagger}c_{i+1}+\text{h.c}+\sum_{i=1}^{L} W_i(n_i-1/2),  
\end{eqnarray}
%
where, $i$ denotes the site index, $L$ the system size, $\thop$ is the hopping amplitude and $U$ is the nearest neighbour interaction strength.  We consider random, uncorrelated on-site potentials, $W_i$, uniformly distributed in the domain $[-W, W]$ and  choose $\thop=1.0$. The filling fraction, $N/L$, is taken to be $1/2$. 

Our methodology for evaluating the numerically exact time evolution is essentially Chebyshev propagation~\cite{Wei06}; details have been explained in Refs.~\onlinecite{Bera2017, Weiner19}. 
We here describe only our  tdHF procedure. 
The time-dependent observables 
corresponding to a time-evolving many-body state $|\Psi(t)\rangle$ can be deduced from the density matrix
$
\varrho_{ij} = \langle \Psi(t)| c_{i}^{\dagger}c_{j}|\Psi(t)\rangle, 
$ 
and the corresponding equation of motion 
\eq{
i \dot{\varrho}_{ij}(t))=\langle\Psi(t)|[c_{i}^{\dagger} c_j, H]|\Psi(t)\rangle. 
\label{Dyn}
}
At this stage,the wave function $|\Psi(t)\rangle$ is arbitrary; it will be specified below by choosing the initial condition for the time-integration of \eqref{Dyn}.

\begin{figure*}[tbh]
    \centering
    {\includegraphics[width=1.0\textwidth]{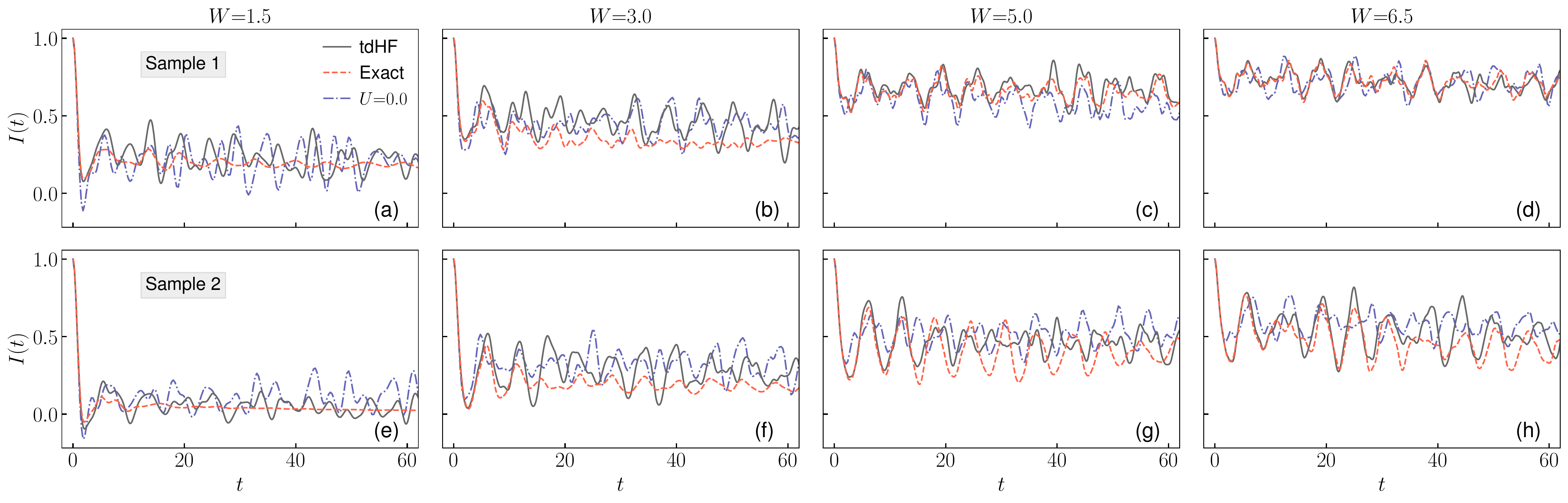}}
    \caption{
     Imbalance $I(t)$ as a function of time for two different samples and four disorder values ($L=24, U=1.0$). The numerically exact time evolution of the interacting system is compared with tdHF and with the non-interacting $I(t)$. The plot illustrates how the decreasing tendency towards equilibration with increasing $W$ can be read off not only from the average imbalance but also from the temporal fluctuations (variances) of $I(t)$. 
     \label{F2}}
\end{figure*}

The Hartree-Fock decoupling of the equation of motion is straightforward~\cite{ringschuck}. We express the resulting time-dependent Hartree Fock (tdHF) dynamics in the (single-particle) eigenfunctions, $\phi_\alpha$, and eigenvalues, $\epsilon_\alpha$, of the non-interacting Hamiltonian $H^{(0)}$: 
\eq{
\label{Dyn2}
i \dot{\trho}_{\alpha \beta}(t) &= i(\epsilon_{\alpha}-\epsilon_{\beta})\trho_{\alpha \beta} \\ \nonumber &+\sum_{\ua\ub\ug}(\trho_{\alpha \ub}\trho_{\ua\ug}-\trho_{\alpha \ug}\trho_{\ua\ub})(U_{\ua\ub \beta \ug}-U_{\beta \ub \ug \ua })) \\ \nonumber
&+(\trho_{\ua \beta}\trho_{\ug\ub}-\trho_{\ua\ub}\trho_{\ug\beta})(U_{\ua \alpha \ug\ub}-U_{\ua\ub\ug \alpha})). 
}
with interaction matrix elements given by
\eq{
U_{\alpha\beta\gamma\delta}=U\sum_{i}\phi_{\alpha}(i)\phi_{\beta}(i)\phi_{\gamma}(i+1)\phi_{\delta}(i+1). 
}
Throughout this paper, we consider the time evolution as it results from a quench of a charge density wave (``N\'eel state'') at time $t=0$. We use a standard Runge-Kutta (RK4) integration routine for the time evolution of the density matrix; further details about discretization and benchmarks  are given in Appendix~\ref{aTDHF}.
Our main observables are the local particle density, 
$n_j(t)\coloneqq \rho_{jj}(t)$
and the imbalance of particles situated at even and odd lattice sites:
\eq{
I(t)\coloneqq 2/L\sum_{j=1}^{L}(-1)^{j} \la n_j(t) \ra. 
\label{eq:II}
}
The latter is frequently studied in numerical and experimental works, because 
its relaxation behavior distinguishes ergodic from non-ergodic phases~\cite{AbaninBloch-Review-2018}.

\section{Results}
%

\subsection{ Individual sample: fluctuations and dephasing} 

Disordered wires tend to exhibit strong fluctuations of observables in space and also between samples that exhibt different disorder configurations. An  illustration is given with Fig.~\ref{F1}, which displays the time evolution of the particle density at four different  wire sites in two disorder realizations. The  sample-to-sample fluctuations of the corresponding global variables - as exemplified by the charge imbalance -  are displayed in Fig.~\ref{F2} for two different samples at four disorder values. Since such variations are washed out when considering ensemble averaged observables, we analyze the time series for individual samples and ensemble averages separately. 
%

\subparagraph{Temporal fluctuations of local density, $n_j(t)$.}
We analyze the temporal fluctuations of $n_j(t)$ at observation sites equally spaced along the ring, see Fig.~\ref{F1} (a)-(h). 
At moderate disorder, $W{=}1.5$,  the exact time evolution exhibits pronounced temporal fluctuations that are efficiently damped by correlation effects. By inspection one infers that if one were to associate a damping rate, $\Gamma_j$, 
with local fluctuations then $\Gamma_j$ would be seen to fluctuate from site to site. 
By comparing the saturated value of the density $n_j(t)$ as obtained with exact dynamics with the equilibrium value (calculated separately, not shown in Fig.~\ref{F1}), we have confirmed that relaxation is indeed against the thermal value. 
Moreover this relaxation is mostly due to correlation effects beyond HF: 
While the time evolution of $n_j(t)$ within tdHF differs from the non-interacting trace, it also deviates from the exact result. In particular, in tdHF we do not observe the strong damping characteristic of the exact trace in Fig.~\ref{F1} (a)-(h).

At larger values of disorder, $W{=}5.0$ in Fig. \ref{F1} (i)-(p),
 the non-interacting localization length $\xi_\text{sp}$ is of the order of the lattice spacing, 
 $\xi_\text{sp}/a \sim 1$.
In this regime, dephasing is seen to be very much reduced with a damping-behavior that shows large spatial fluctuations. 
In regions with very weak dephasing, tdHF follows the exact trace closely, becoming quantitative in the window of observation times. 

In Fig. \ref{F1} (k) and (o) the interacting traces (exact and tdHF) exhibit very pronounced oscillations that differ in amplitude and frequency from the non-interacting reference revealing a many-body character.  
%
Within the window of observation times shown in this plot, there is hardly any dephasing discernible. The origin of these oscillations we tentatively assign to cooperative effects in a (largely) decoupled two-particle system.
They are very long lived and therefore are an important manifestation of a lack of ergodicity. 


\subparagraph{Imbalance fluctuations, $I(t)$.}
Fig.~\ref{F2} shows the imbalance after a quench from a N\'eel state in two typical samples, for four values of the disorder. Note that for a finite size sample, $I_\text{eq}{\coloneqq}\lim_{t\to\infty}I(t)$ will differ from zero even at weak disorder, i.e. in the thermal regime; 
instead, sample-to-sample fluctuations are expected with a mean value $\oI_\text{eq}$  that vanishes and the corresponding variance $\overline{(I_\text{eq}{-}\oI_\text{eq})^2} \sim L^{-1}$; see Ref.~\onlinecite{Doggen2018} for a similar conclusion.
Therefore, the exact traces shown 
at $W=1.5,3.0$ 
(Fig.~\ref{F2}(a),(b),(e) and (f)) do not tend towards zero at large times even though the system is expected to equilibrate, eventually. 

Not surprisingly, we witness in $I(t)$, Fig.~\ref{F2}, the same qualitative behavior already seen with $n_j(t)$:
at  stronger disorder the signatures of oscillations in $n_j(t)$ carry over to $I(t)$.
Note that these oscillations will be washed out by the (incoherent) spatial  averaging in large samples
that is the defining feature of global variables. 
Thus the local, only weakly damped  temporal fluctuations are captured by global observables only as a finite-size effect. 
For the same reason they also are eliminated in ensemble averages, as we show in the next subsection. 

\begin{figure}[t]
    \centering
   {\includegraphics[width=1.0\columnwidth]{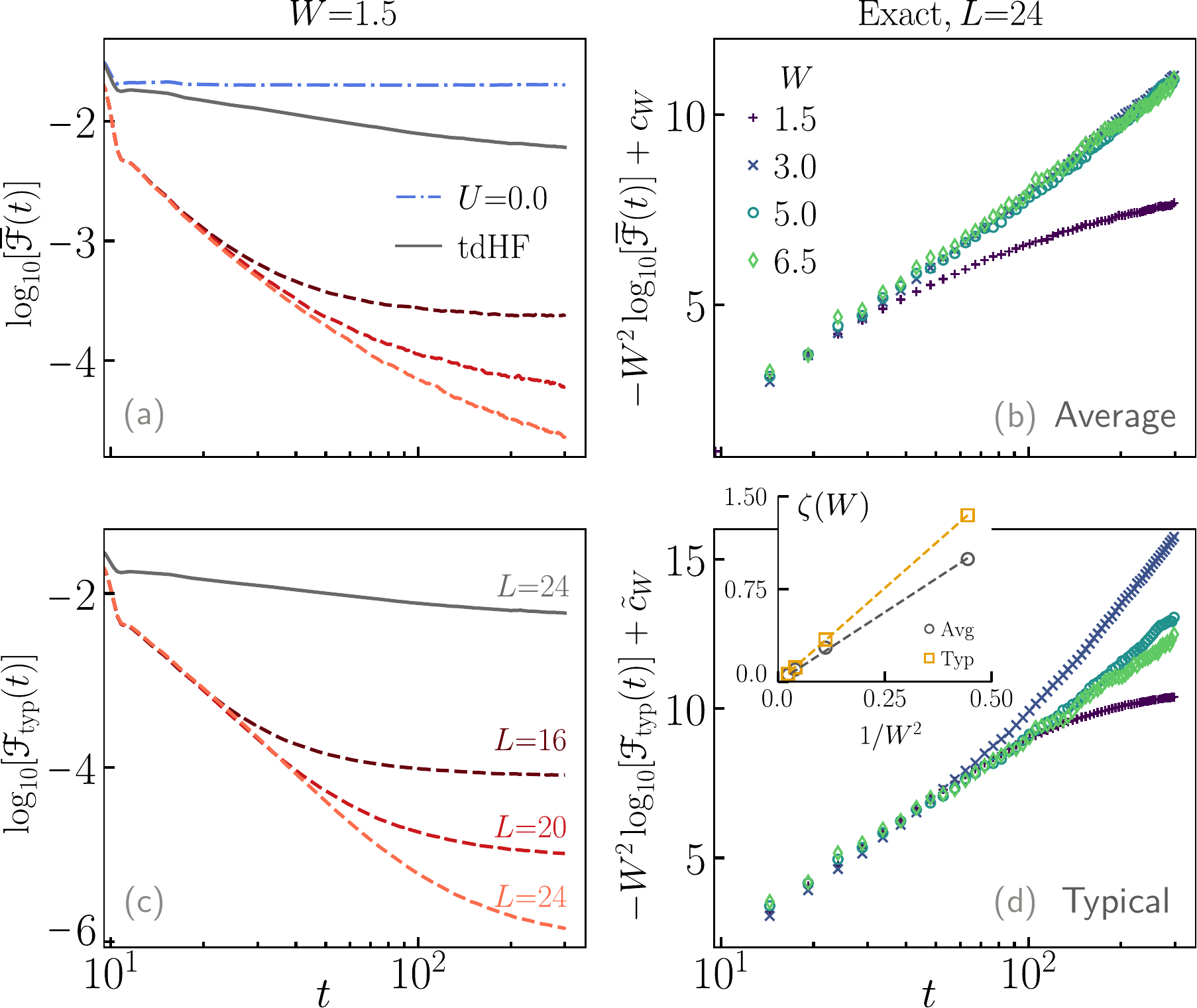}}
    \caption{
     Local time fluctuations of disordered average and typical $\msr{F}(t)$ as a function of time. (left column, panels (a),(c)) A power-law decay of the tdHF and exact traces at disorder strength $W{=}1.5$ is observed in average and typical fluctuations for different system size $L=\{16, 20, 24\}$. The non-interacting (blue) trace in Fig. \ref{fig:timefluc}(a) shows no decay as expected. The right column (Fig. \ref{fig:timefluc}(b),(d)) highlights the power-law scaling with an exponent which is roughly proportional to $\zeta(W)\sim W^{-2}$; $c_W$ is a non-universal prefactor that depends on $W$. The inset displays the fitted exponents.  }
     \label{fig:timefluc} 
\end{figure}

\subsection{ Ensemble averages: temporal fluctuations and imbalance} 
%
\subparagraph{Temporal fluctuations of $n_j(t)$.}
We quantify the temporal fluctuations of local charge density by the ``running variance'' per sample 
\eq{
\msr{F}(t) = 1/L \sum_{j=1}^L \la [n_j(t) - \la n_j(t) \ra_{\Delta t}]^2 \ra_{\Delta t},
\label{eq:fluc}
}
where $\la \ra_{\Delta t}$ denotes a sliding time window 
average~\footnote{{The moving average for a function $f(t)$ is defined as 
$\langle f(t) \rangle_{\Delta t} \coloneqq (\Delta t)^{-1}\int_{t-\Delta t/2}^{t+\Delta t/2} f(t^\prime) dt^\prime$ for practical calculations we adopt the discretized version. An analogous observable has also been investigated in the context of trapped ion simulators~\onlinecite{KaplanTF20}.}};
here an averaging window $\Delta t \approx 10 $ was chosen wide enough 
for a few oscillations in $n_j(t)$ to fall 
within, see Fig.~\ref{F1}. The qualitative results are largely insensitive to the specific numerical choice made here. 

Figure~\ref{fig:timefluc}(a,b) displays the disorder averaged sample variance, $\bar{\msr{F}}(t)$, 
while the lower panel (c,d) shows the corresponding typical fluctuations 
$\msr{F}_{\text{typ}}(t) \sim \exp(\overline{ \log {\msr{F}(t)}})$, 
where again the overline denotes ensemble averaging.

As demonstrated in Fig. \ref{fig:timefluc}, damping of temporal fluctuations is not exponential; instead, 
a wide time window exists with (approximate) power-law decay of the average variance, $\sim t^{-\zeta}$, 
featuring a non-universal (possibly effective) exponent $\zeta(W)$. The numerical estimates are shown in the inset of Fig. \ref{fig:timefluc}; for instance, at moderate disorder, $W=1.5$, we observe $\zeta$ of order unity. 
As compared to  averages, the associated typical observables, 
$\msr{F}_{\text{typ}}(t)$, exhibit a faster decay with a larger 
exponent $\zeta_{\text{typ}} \sim 1.34$, which reflects very large sample-to-sample fluctuations. 
Note that  the exponent extracted from exact traces 
exceeds the corresponding tdHF value, $\zeta\sim 0.17$, considerably, indicating the correlated character of the damping mechanism. A similar information is reflected in the deviation of  average and typical fluctuations, being less than 2\% for tdHF as compared to an order of magnitude for the exact traces, see Fig.~\ref{fig:timefluc}(a,c).  

{\it Discussion:} (i) Fig.~\ref{fig:timefluc}(a,c) exhibit very strong finite size effects; in particular, the typical traces displayed in 
Fig.~\ref{fig:timefluc}(c) exhibit a small curvature indicating a flow to effective exponents $\zeta(W)$ that potentially 
grow in time. Conceivably, the flow is indicating an asymptotic decay that is exponential with an asymptotic 
rate, $\Gamma(W)$, which vanishes as $W$ approaches $W_\text{c}$ from below; such  a 
scenario is foreseen in Ref.~\onlinecite{Serbyn2014}. According to these authors, damping in the localized phase 
$W>W_\text{c}$ is described by a power law, $t^{-2b}$, with $2b\sim \xi_\text{sp}$. 
It is tempting to associate $2b$ (defined in Ref. \onlinecite{Serbyn2014} at $W{>}W_\text{c}$, and also observed in strongly disordered phase within a  perturbative calculation in Ref.~\onlinecite{detomasi19}.) 
with $\zeta$ (observed here also at $W{\lesssim}W_\text{c}$) 
by assuming that $\zeta$ describes a pre-asymptotic regime where damping proceeds
invoking the same microscopic mechanism prevalent also in the localized phase. 
Under these premises one might try $\zeta \sim W^{-2}$, 
since $\xi_\text{sp}\sim W^{-2}$ in a regime $\xi_\text{sp}\lesssim a$. At first sight such a scaling is indeed compatible with our numerical data, 
see inset Fig.~\ref{fig:timefluc}(d). Nevertheless, it seems premature to identify $\zeta$ with $2b$ at this 
stage; for instance, this would imply that -- within the same pre-asymptotic time window -- 
at $W<W_\text{c}$ damping proceeds as in the localized phase, while 
simultaneously the density propagator already exhibits (sub-)diffusive behavior. 
Such a coexistence of dynamical behavior would certainly merit extra attention.

(ii) Dephasing as observed for $n_j(t)$ leaves a trace also in the running variance of imbalances defined as 
$
\msr{F}^{(I)}(t) = \la [I(t) - \la I(t) \ra_{\Delta t}]^2 \ra_{\Delta t}$. 
The corresponding analysis has been relegated to the appendix, Sec. \ref{appimb}.

\subparagraph{Averaged imbalance.}

\begin{figure}[t]
    \centering
    {\includegraphics[width=1.0\columnwidth]{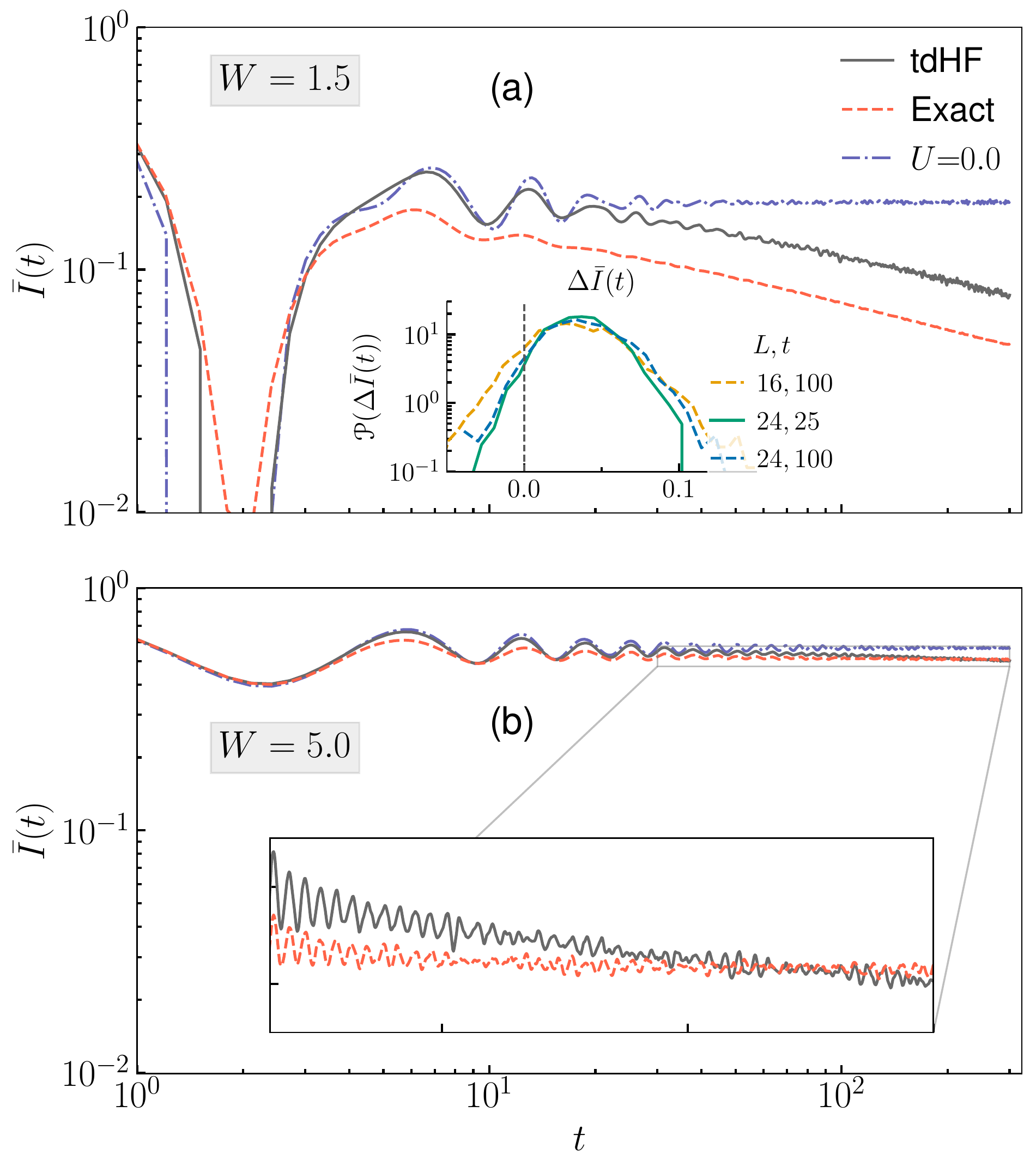}} 
    \caption{Exact imbalance($\oI(t)$: red, dashed) and tdHF-result (black, solid)  as a function of time  for two different values of disorder strengths $W{=}{1.5, 5.0}$ (parameters: $U{=}1.0, L{=}24$,  averaged over $\sim 1000$ samples).  
    For comparison the non-interacting trace is also shown (blue, dot-dashed). Inset (a) shows the distribution of time averaged deviation $\msr{P}(\Delta \overline{I}(t))$ between the exact and tdHF trace at two different times highlighting that the exact trace falls below the tdHF result, typically, at shorter times. 
    Inset (b) shows the decay of the tdHF-imbalance at long times where the exact $\oI(t)$ nearly 
    saturates at this system size, signalizing ergodicity breaking at large $W$. 
    \label{fig:avgImb}}
\end{figure}

Figure~\ref{fig:avgImb} shows the ensemble averaged imbalance, $\oI(t)$, in the ergodic phase $W{=}1.5$ and also at strong disorder,  $W{=}5.0$. Pronounced oscillations are seen in these traces. They occur already in the absence of interactions, see Fig. \ref{fig:avgImb}, and therefore are not related to the cooperative oscillations and dephasing discussed before; as expected, the strong temporal fluctuations seen in single-sample traces have averaged out, 
so information about dephasing rates has been eliminated. Concerning the exact averaged trace, 
$\oI(t)$,  one is left with the well known observation that it vanishes  at large times, 
$\oI(t) \sim t^{-\beta(W)}$, where  $\beta {=} 1/2$ in a diffusive system~\cite{BarLevPRL2015, Luitz2016, Bera2015}.

Concerning the numerical estimate of the exponent $\beta$, several works pointed out that finite size effects are strong and therefore the asymptotic regime is very challenging to reach; correspondingly, the observed exponents 
could be effective in the sense that they approximate pre-asymptotic, transient behavior~\cite{Bera2017, Doggen2018, Schulz_2020, PandaMBL19}. 
Specifically, at finite system sizes with restricted time window a smaller  than $1/2$ exponent is observed in exact traces even at moderate disorder; slowing down of the dynamics has been associated with rare region effects~\cite{Vosk2015, GopalakrishnanPRB15, AgarwalPRL2015, Luitz2016,Luitz:2017cp, Agarwal2017,KnapHFPRB18, Doggen2018}. 
However, since slowing down has been observed recently also in the Aubry-And\'re model that is unsuspected of exhibiting rare regions, this interpretation is challenged~\cite{Weiner19}.
\begin{table}[t]
\begin{tabular}{p{1.25cm}|ccc|ccc}
    \hline \hline 
     \multirow{1}{*}{Method} &
      \multicolumn{3}{c|}{$W=1.5$} &
      \multicolumn{3}{c}{$W=3.0$} \\
        \cline{2-7}
 & $\beta$ & $L$ & $t~[\thop^{-1}]$ & $\beta$\ & $L$ & $t~[\thop^{-1}]$   \\
\hline 
Exact & ${\sim}0.21$ & $[16{-}24]$ & $[12,60]$  & ${\sim}0.07$ & $[16{-}24]$ & $[65,150]$ \\
tdHF  & ${\sim}0.33$  & $[16{-}32]$ &$[50,300]$ &${\sim}0.15$  & $[16{-}32]$ & $[100,300]$\\
Luitz et. al.~\cite{Luitz2016} & ${\sim}0.2$ & $[16{-}24]$ & - & ${\sim}0.05$ & $[16{-}24]$ & -\\
Doggen et. al.~\cite{Doggen2018} & - &  &  & $\sim 0.07$ & $[50{-}100]$ &  $[50, 100]$ \\
\hline
\end{tabular} \label{tab}
\caption{Comparison of the flowing exponent $\beta(W, L)$ in the pre-asymptotic regime extracted from both exact and tdHF imbalance $\oI(t)$. Here we also compare with some existing results of the exponent in the ergodic phase. Note, for comparison the above exponent from exact traces $\beta$~(first row) is calculated for the time window where the system sizes overlap, however, with increasing $L$ the time window increases and usually allows the exponent to flow towards a higher value~\cite{Bera2017}, which is ignored here. }
\end{table}
To be more quantitative we show the effective exponent $\beta(W,L)$ in Tab.~\ref{tab} for different values of $W$. 

\begin{figure}[b]
    \centering
    \includegraphics[width=1.0\columnwidth]{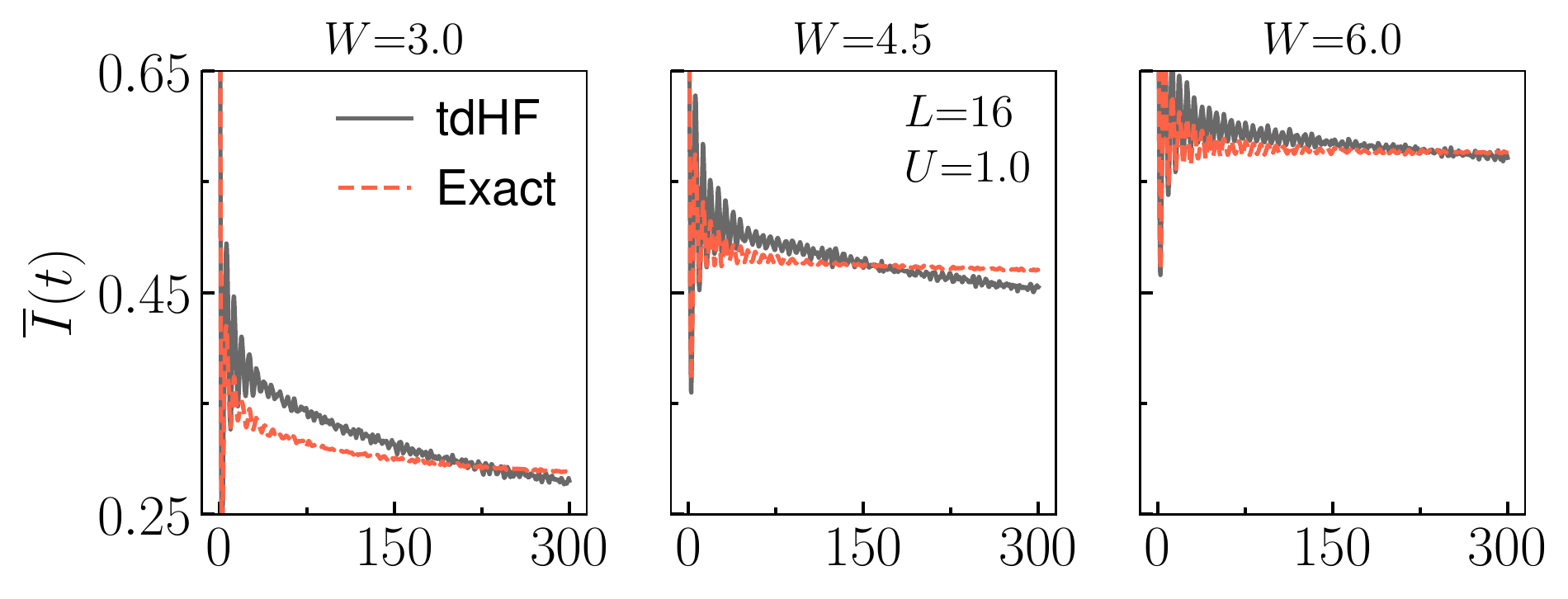}
    \caption{Traces similar to the previous Fig.~\ref{fig:avgImb} at smaller system size $L=16$ and at $W{=}3.0, 4.5, 6.0$. Data demonstrates the existence on an intersection point highlighting the two different time regimes in exact traces: short time/strong dephasing and long time/localizing trend. } 
    \label{f4}
\end{figure}

{\it Discussion.} As one would expect at weak disorder, it is seen in Fig.~\ref{fig:avgImb}(a) that 
the tdHF-trace follows the non-interacting one a bit longer than the exact one 
before it deviates towards lower values. At larger times similar to the exact trace, 
also the tdHF-dynamics tends to thermalize albeit with 
a different (sliding) exponent, see Tab.~\ref{tab}.
\footnote{We verified by inspecting several samples that  the exact trace falling below the tdHF one 
is indeed typical for short observation times, 
being more pronounced for $W{=}3.0$~(see Fig.~\ref{F2}(b,f)).
To further illustrate this point, 
the inset of Fig.~\ref{fig:avgImb}(a) shows the distribution of the integrated deviation 
$
\Delta I(t) {\coloneqq} 1/t \int_{0}^t dt^\prime (I_{\text{tdHF}}(t^\prime) {-}  I_{\text{exact}}(t^\prime) )
$
at two times, $t{=}\{25,100\}$. 
The main weight of the distribution  $\msr{P}(\Delta I(t) )$ is seen to be positive; 
its width shrinks with increasing system size $L$.  }
Note that the tdHF exponent, $\beta_\text{tdHF}$, is consistently larger than 
the exponent that is found from the exact $\oI(t)$ in the ergodic phase. 
As a result one expects the tdHF-trace to intersect with the exact one at large times. 
This point is illustrated in the inset of Fig.~\ref{fig:avgImb}(b) which confronts the evolution of  
$\oI(t)$  with the corresponding tdHF result at large disorder, $W{=}5\thop$. 
It is seen that at large times the exact trace being nearly horizontal displays a localization phenomenon. 
Its many-body character reveals from the fact that the tdHF-trace intersects the exact one and falls below. 
This is indicating a tdHF-tendency towards thermalization which is absent in the exact evolution~\footnote{Note, that a similar decay of self-consistent-field traces 
can be detected in the data of Ref.~\onlinecite{KnapHFPRB18},
where it was discussed in the context of self-consistent noise.} 
As seen from Fig.~\ref{f4}, such an intersection point is  encountered also at other disorder values, 
especially weaker ones,  which suggests that it occurs generically. 

\section{Heuristic argument linking dephasing and multifractality}
We present a heuristic argument that explains the strong dependency of $\zeta(W)$ on $W$, by connecting the exponent with the non-interacting (possibly renormalized) localization length $\xi_\text{sp}$ and an exponent $\alpha(W)$ signalizing multifractality: $\zeta(W)\approx \alpha(W)(\xi_\text{sp}/a)$. 

Our consideration starts with an argument based on wavefunction overlap:  the {\it bare}  coupling $J$ of the charge degree of freedom  at the origin to the charge dynamics a distance $x$ away from the origin is exponentially small, $J\approx \thop^* e^{-x/\xi_\text{sp}}$, where $\thop^*$ denotes an effective coupling inside the localization volume. Charge correlations establish over the distance $x$ if the action $tJ$ becomes of order unity;  hence, we are led to define a typical correlation time $t\approx  e^{x/\xi_\text{sp}}/\thop^{*}$. Conversely, for a fixed time we can define a correlation volume $x(t)\approx \xi_\text{sp}\ln(t\thop^*)$.

We will now consider a quench from an initializing  many-body state $|\Psi\rangle$. 
We then have for the dynamics after a quench, $t>0$, the exact expression: 
$$ n_0(t)=  \sum_{\alpha\beta} n_{\alpha\beta} f_{\beta\alpha} e^{-\mathfrak{i}(E_{\alpha}-E_\beta)t} $$
with a weight  
$f_{\beta\alpha}{\coloneqq}\langle \Psi|\alpha\rangle\langle\beta|\Psi\rangle$ and a matrix element  $n_{\alpha\beta}{\coloneqq}  \langle\alpha|\hat n_0|\beta\rangle$. 

The sum is over all eigenstates, $|\alpha\rangle,|\beta\rangle$, of the many-body Hamiltonian of the full system. 
We now interpret the concept of the correlation space as implying an approximate representation
\begin{equation} 
n_0(t)\approx  \sum^{\mathcal{H}_\text{corr}(t)}_{\alpha'\beta'} 
\mathfrak{n}_{\alpha'\beta'}
\mathfrak{f}_{\beta'\alpha'} 
e^{-\mathfrak{i}(E'_{\alpha'}-E'_{\beta'})t}; 
\label{eB1} 
\end{equation}
here, the sum is over the Hilbert space of the correlation volume, $\mathcal{H}_\text{corr}(t)$, and the energies $E'$ denote the  quasi-energies of the Hamiltonian projected on the many-body states of the correlation volume.

\begin{figure}[tbh]
    \centering
   {\includegraphics[width=1.0\columnwidth]{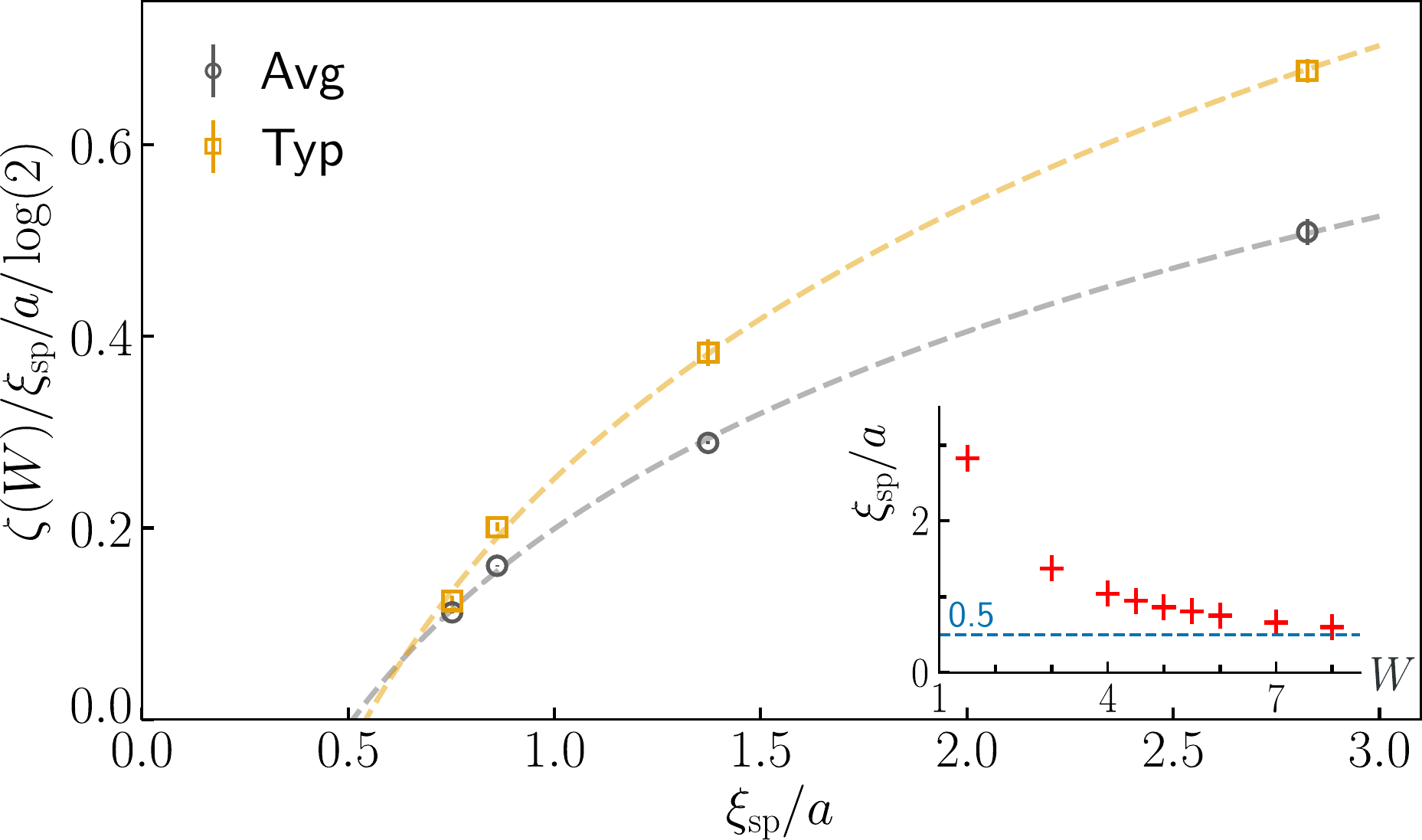}}
   \caption{\label{f6} The exponent $\zeta(W)$ as obtained from Fig. \ref{fig:timefluc}(b),(d) plotted over the non-interacting localization length $\xi_\text{sp}(W)$. 
   (data points represent disorder: $W{=}1.5,3.0,5.0,6.5$). The plot highlights the residual dependency, $\alpha(W){=} \zeta(W)/(\ln (\lambda)  \xi_\text{sp}/a)$ with $\lambda{=}2$, which has the interpretation of the fractal dimension of the dynamically active fraction of the many-body Hilbert space. Dashed lines are guides to the eye. Inset shows the $W$ dependence of extracted $\xi_{\text{sp}}$ from the infinite temperature many-body density-density correlator as described in Ref.~\onlinecite{Weiner19}.}
\end{figure}

\newcommand{\cD}{\msr D}
We now introduce the dimension $\cD(t)$ of 
${\mathcal H}_\text{corr}(t)$; with $\lambda$ the size of the Hilbert-space per unit length $a$ we have 
$a \ln \cD(t)\approx x(t)\ln(\lambda)$. Then we obtain
 for the amplitudes a scaling  $\mathfrak{f}_{\beta\alpha}\sim \cD^{-1}(t)$ reflecting the normalization of the wavefunction. Further, if we choose the initializing state $|\Psi\rangle$ as an eigenstate of $\hat n_0$ with unity occupation, then we also have $\mathfrak{n}_{\alpha\beta}\sim \cD^{-1}(t)$.  

The concept of the local Hilbert space becomes effective, after taking the coefficients  $\mathfrak{f}_{\beta\alpha}, \mathfrak{n}_{\alpha\beta}$  structure-less. Then, recalling that in the correlation volume by definition the energy difference between two states exceeds $J$, we can stipulate that the exponential in \eqref{eB1} is distributed "randomly" on the unit circle. 
With this idea, the sum \eqref{eB1} can be evaluated by 
assuming that each constituting term is uncorrelated from all the others.

Correspondingly, 
$n_0(t) \sim \cD^{-2+1}(t)$ and thus
\begin{equation}
n_0(t) \sim 
t^{-\zeta_0}, \qquad \zeta_0 =   (\xi_\text{sp}/a)
\ln\lambda 
\end{equation}
The approximate treatment just proposed ignores correlations between coefficients and energies and in this sense is similar to a factorization approximation. As such it is, presumably, uncontrolled. 
An improved approximation will account, e.g., for the possibility of an effective Hilbert space, i.e. dominating multifractal substructures in the full Hilbert space of the correlation volume. A notion of multifractality suggests a replacement $\lambda^{x(t)}\to \lambda^{\alpha x(t)}$ when estimating the effective Hilbert space dimension and a corresponding improved estimate 
\begin{equation}    
\zeta 
= \alpha(W)  (\xi_\text{sp}/a)\ln\lambda, 
\quad 0<\alpha(W)\leq 1.
\label{eq10} 
\end{equation} 
To illustrate and quantify this relation, $\alpha(W)$ has been plotted in Fig. \ref{f6}. The incident of  $\alpha(W)\ll 1$  at small $\xi_\text{sp}$ (i.e. large $W$) reflects the strong  multifractality of the dynamically active many-body Hilbert space. The data is consistent with a freezing transition, $\alpha(W_\text{freeze}){=}0$, taking place at  $\xi_\text{sp}\approx0.5 a$,
which corresponds to $W_\text{freeze}\gtrsim 10$. 
Freezing as it manifests here for a physical observable has been discussed before in the context of many-body wavefunction statistics, e.g., in Ref. \onlinecite{Serbyn2017}.

One would expect the estimate \eqref{eq10} to reproduce the qualitative behavior at intermediate times. This expectation is certainly confirmed by our simulation results. Moreover, the scrambling of information as implied by taking coefficients structureless amounts to an unprejudiced involvement of all of the available Hilbert space ${\mathcal H}_\text{corr}(t)$. Scrambling thus incorporates strong many-body correlations which are beyond mean-field dynamics as resembled with tdHF. This observation explains why our simulations detect a qualitative difference between dephasing as seen in the tdHF-traces from the exact results.  

The heuristic reasoning here presented has been partially inspired by \textcite{Serbyn2014}. Note, however, that these authors have made their case for the many-body localized regime, while our argument relies on (quasi-)ergodicity in the (growing in time) correlation volume and hence is more suitable for the ergodic or critical regimes at moderate disorder. This regime is known to exhibit a wide time window with transient behavior~\cite{Bera2017}, whose parametric boarders are not well known~\cite{Weiner19}. The power-law dynamics we see in ${\mathcal F}$ may reflect an intermediate behavior, which ultimately converges towards an  exponential form; such an evolution would be consistent with Ref.~\onlinecite{Serbyn2014}. 
Note further that \textcite{Serbyn2014} do not include multifractality into their argument, effectively letting $\alpha{=}1$. In hindsight, this appears to be an oversimplification since in the MBL-regime  multifractality tends to be  strong~\cite{MonthusMF_2016, Serbyn2017,AletMFPRL19,pietracaprina2019hilbert, LuitzMF_2020,  tikhonov2020eigenstate}.

\section{Conclusion and Outlook}
We have presented an analysis of the dynamical fluctuations of the local charge density $n_j(t)$ in strongly disordered interacting quantum wires. After a quench from a N\'eel state, $n_j(t)$ exhibits strong temporal  fluctuations that gradually decrease within the observation time. These fluctuations can also be seen in global variables, such as the charge imbalance, $I(t)$, where they manifest as dynamical  finite-size effects in ensemble averaged variances. 
The time decay of such variances is described within our window of observation times by a 
(potentially effective) power law, $t^{-\zeta}$, with an exponent continuously varying with disorder strength, $\zeta(W)$. 
At moderate disorder strength $W$, the exponent is different for average and typical variances reflecting very large sample to sample fluctuations.
While for average variances a power law has been predicted in the localized phase~\cite{Serbyn2014}, we here find it also in parameter regions below the putative many-body localization transition. 

In order to analyze correlation effects, we have compared time traces for numerically exact computations with traces obtained within the time-dependent Hartree-Fock approximation. While tdHF turns out to be a useful diagnostic tool of correlation effects, it misses elementary qualitative physics, in particular the many-body localization: at long times tdHF always exhibits a trend towards delocalization irrespective of the regime of disorder. 
We assign this trend to temporal fluctuations in the self-consistent field. These conclusions are at variance with 
claims made previously in Ref.~\onlinecite{KnapHFPRB18}. 
While finishing the manuscript we became aware of closely related work Ref.~\onlinecite{Mirlin2020}, which arrives at similar  results.

Our results have implication for experiments on cold atoms. Indeed, imbalances have already been obtained in experiments at observation times and system sizes comparable to our numerical study. We propose to analyze the measured observables, in particular the per-sample imbalance, in terms of ensemble-averaged time-dependent variances. 
Based on our study, a power-law regime - possibly transient - should be found with exponents $\zeta$ that depend on disorder strength. For purely random potentials a dependency $\zeta =\alpha(W)\ln(\lambda) (\xi_\text{sp}/a)$ is predicted, with $\xi_\text{sp}$  being the non-interacting localization length and $\alpha(W)$ a multifractal scaling index representing the volume fraction of the dynamically active Hilbert space within the full Hilbert space. A computational study of the (experimentally  relevant) Aubry-Andre-model is  currently under way.

\section{ACKNOWLEDGMENTS}
We would like to thank G. De Tomasi, E. Dogger, J. Karcher, A. D. Mirlin, P. P\"opperl, K. Richter, R. Sensarma, K. Tikhonov and J.-D. Urbina for several discussions and useful comments on the manuscript. 
SB acknowledges support from SERB-DST, India, through Ramanujan Fellowship (No. SB/S2/RJN-128/2016), Early Career Research Award (No. ECR/2018/000876), Matrics (No. MTR/2019/000566), and MPG for funding through the Max Planck Partner Group at IITB.
SN would also like to thank the MPI-Partner group program for financial support. Support from German Research Foundation (DFG) through the Collaborative Research Center, Project ID 314695032 SFB 1277 (projects A03, B01) and through EV30/11-1, EV30/12-1 and EV30/14-1 are  acknowledged. 

\bibliography{MBL}

\appendix 
\renewcommand\thefigure{\thesection\arabic{figure}}    

\begin{figure*}[thp]
    \centering
   {\includegraphics[width=1\textwidth]{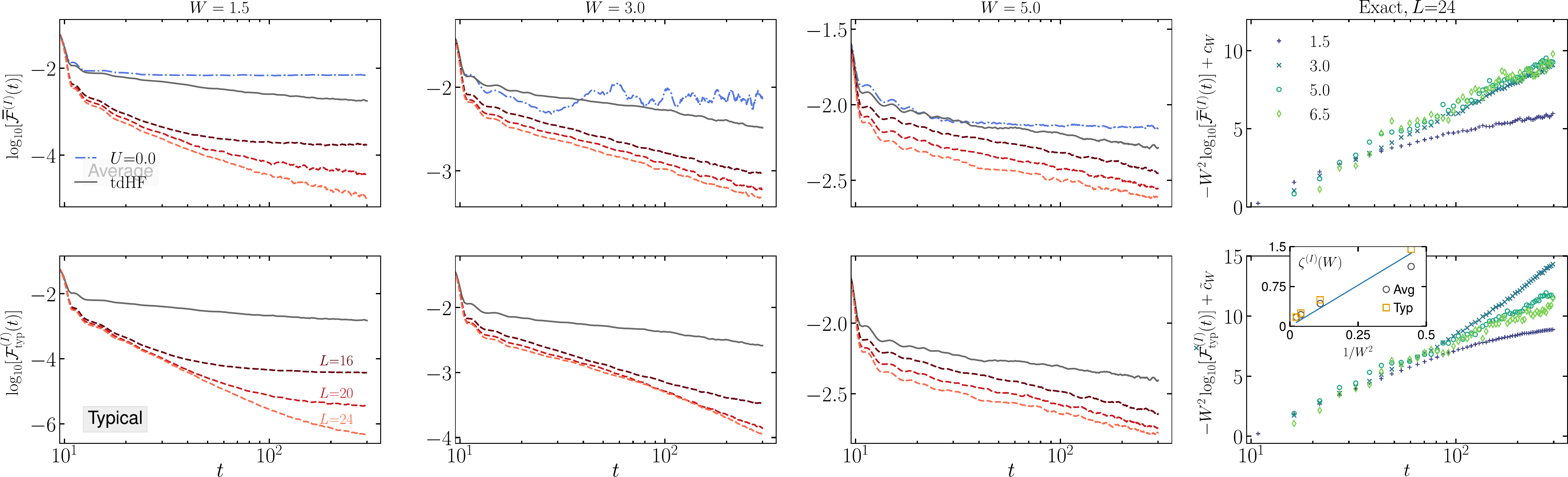}}
    \caption{
     Time fluctuations $\msr{F}^{(I)}(t)$ of (running variance) of the imbalance $I(t)$ taken over the ensemble as average value  (upper panel) and typical value  $\msr{F}^{(I)}(t)$ (lower panel), here plotted as a function of time. (Parameters: interaction $U={1.0}$; disorder $W=1.5, 3.0, 5.0$; system size $L=\{16, 20, 24\}$.) The non-interacting trace shows no decay as expected. Rhs column shows an attempted collapse of exact data for $L=24$ with an exponent $\zeta^{(I)}(W) \sim \xi_{\text{sp}}$. Inset:  the (effective) exponent $\zeta^{(I)}(W)$ roughly estimated from systems with $L=24$ and time intervals $[20,50]$. 
     \label{fig:timeflucimb}
     }
\end{figure*}

\section{Time fluctuation of imbalance}
\label{appimb}
Figure~\ref{fig:timeflucimb} shows a similar analysis of the statistics of the local time fluctuation 
$$
\msr{F}^{(I)}(t) \coloneqq \la [I(t) - \la I(t) \ra_{\Delta t}]^2 \ra_{\Delta t}. 
$$ (per sample) associated with $I(t)$ as evaluated previously for the local density $n_j(t)$ in the main text~(see Fig.~\ref{fig:timefluc}). 
Also for the fluctuations of the imbalance $I(t)$ we observe strong finite size effects, which become more severe with increasing disorder in both average and typical traces.

Nevertheless, as was the case for $n_j(t)$, 
also for the averaged fluctuations $\msr{F}^{(I)}(t)$ we advocate a (possibly transient) power-law decay, $t^{-\zeta^{(I)}}$, being more prominent in the typical traces. Following this observation, the traces for the largest system size exhibit an (approximate) scaling collapse with $\zeta^{(I)}(W) \propto 1/W^2$.
The overall behavior obtained for the damping of the fluctuations of the density and the imbalance is thus seen to be qualitatively the same, matching expectations.

\section{Further assessment of tdHF in the context of MBL} 

We reiterate the results reported in the main text: 
At weak disorder, the strong dephasing seen in exact traces is not reproduced by tdHF, which exhibits a much weaker damping rate. We take this as an indication that the energy and momentum exchange mediated by time-dependent mean-fields is a rather weak damping mechanism as compared to the two-particle scattering events contained in correlation effects. Quantitatively this manifest itself as faster decay of traces for tdHF compare to the exact traces~(see Tab.~\ref{tab}).  

At strong disorder, tdHF predicts thermalization while a (nearly) localized dynamics is seen in exact calculations. 
This difference, i.e. the ``absence of thermalization'' in exact traces as compared to tdHF, we interpret it as a strong signature of  MBL physics. 
One way to interpret this result would be to assign it to the absence of many-body quantum interference in tdHF but, presumably, simpler effects factor in as well. For example, charge fluctuations at large disorder will be suppressed by effects related to the Coulomb-blockade, which are also not accounted for in tdHF.

We conclude that tdHF-dynamics deviates qualitatively from the exact time evolution. As a diagnostic tool with respect to MBL-physics and correlations, tdHF has its merrits. 
 However, tdHF-traces have a tendency towards equilibration even at large disorder indicating that MBL-physics is not appropriately included. 
In the next subsections we give further evidence of our claims.

\subsection{System size dependence of tdHF simulations}
\label{ltdhf}

\begin{figure}[t]
    \centering
    {\includegraphics[width=1.0\columnwidth]{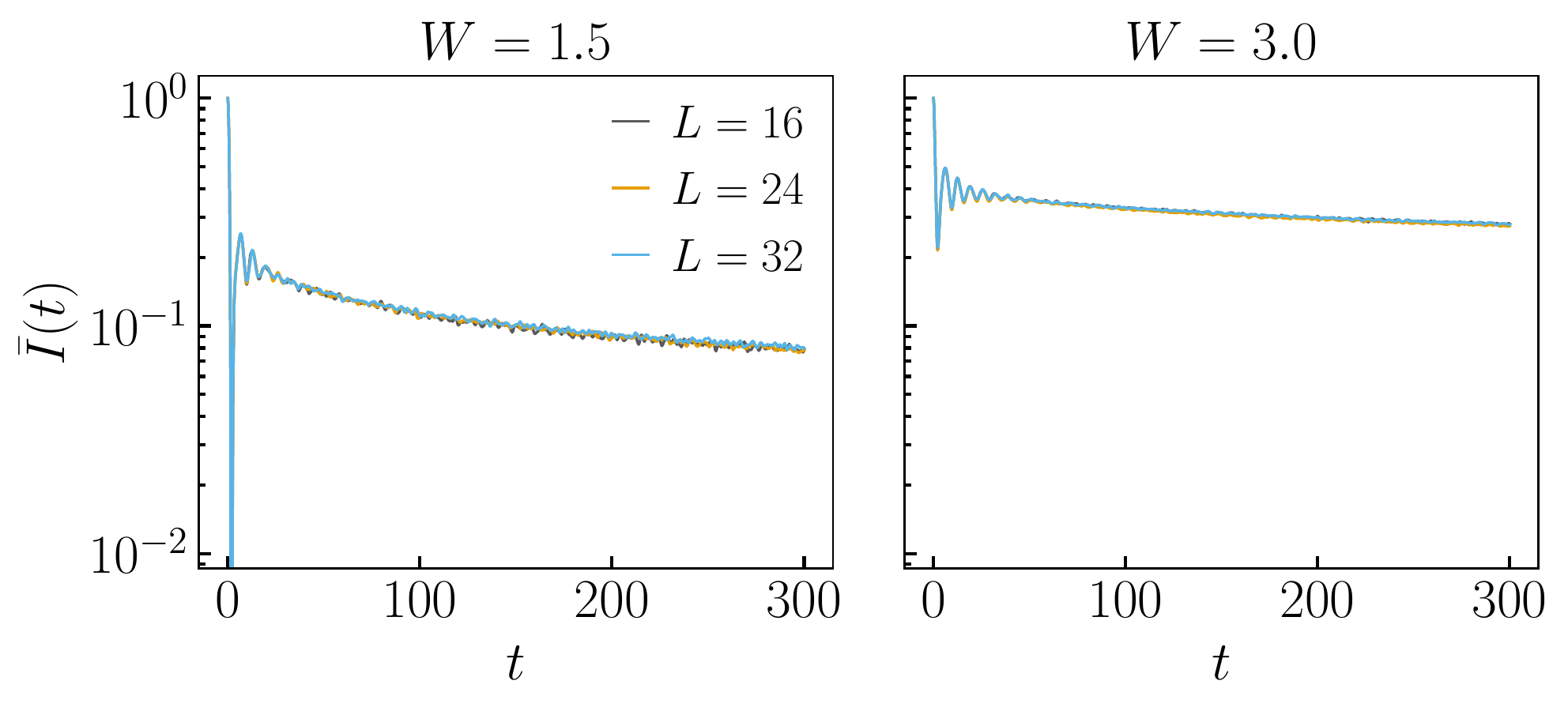} }
    \caption{ Average imbalance $\oI(t)$ as obtained with tdHF. The plot illustrates the system size insensitivity of the tdHF for disorder values $W=1.5, 3.0$ and $U=1.0$. The data is averaged over $\sim 1000$ disorder configurations. 
        \label{tdhfL}}
\end{figure} 
A further qualitative difference manifesting in tdHF time evolution as compared to the exact dynamics reveals in finite size effect. 
Figure~\ref{tdhfL} shows the system size dependence of tdHF $\oI(t)$ for two different values of disorder strength. Within our simulation time ($t \lesssim 300 \thop$) we do not observe any significant dependence on $L$ in the tdHF traces, very much in contrast to what we observe for the exact traces shown in Fig.~\ref{fig:timefluc}. This further
underlines the significance of correlations and - in particular - their important role for the finite-size effects on the density relaxation. Their ubiquitous and pronounced appearance in the MBL-problem remains to be understood.

\begin{figure}[t]
    \centering
    {\includegraphics[width=0.9\columnwidth]{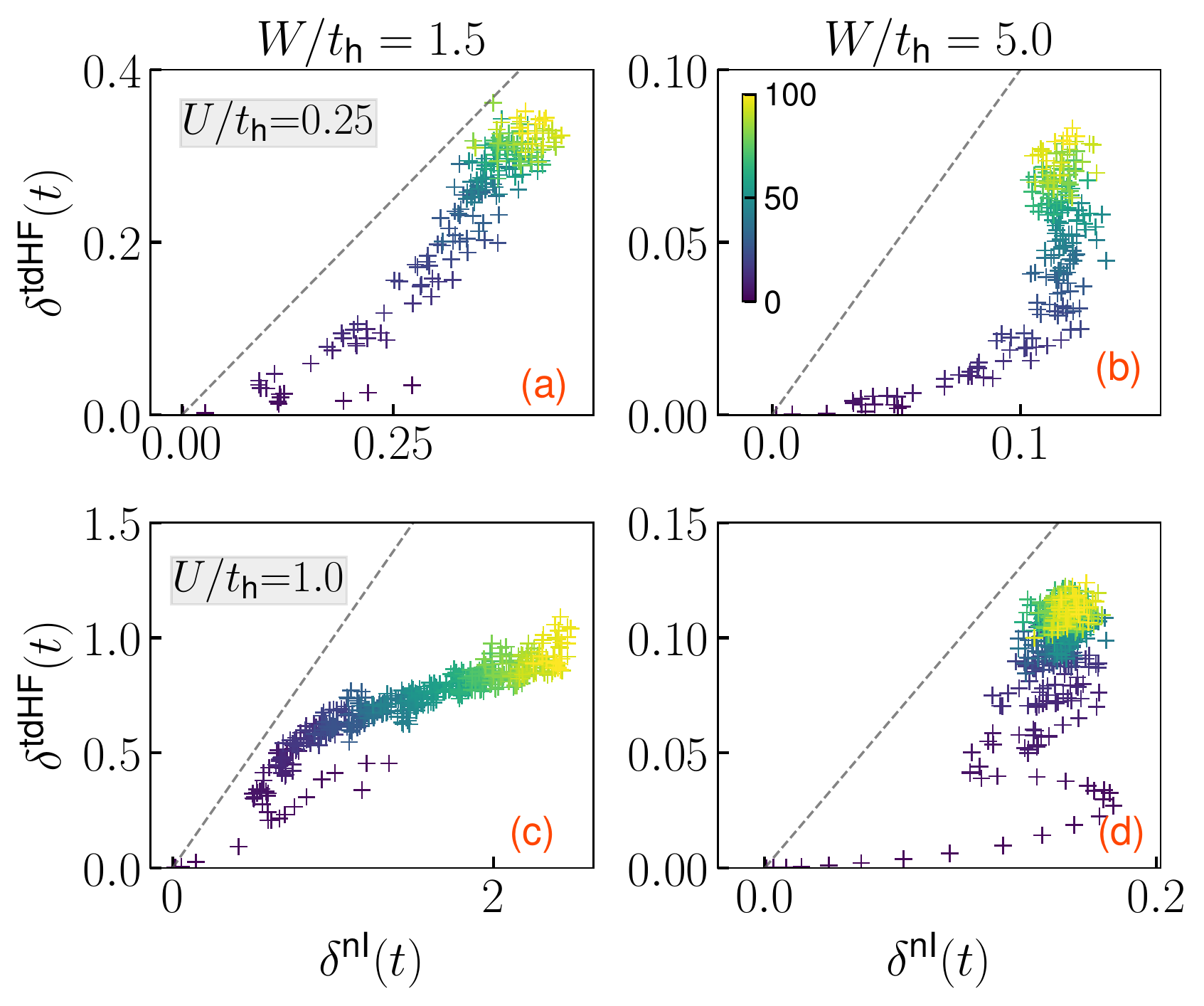}} 
    \caption{ Deviation $\err{tdHF}$~(see Eq.~\eqref{eq:err}) shown as a function of $\err{nI}$ for different values of interaction~($U=0.25, 1.0$ per row) and disorder strengths~($W=1.5, 5.0$ per column). The inset shows the colorbar for time argument, longer time implies lighter color. The data is shown for $L=24$ with $500$ disorder configurations. The dashed line indicates unity slope indicating 
    $\err{tdHF}{=}\err{nI}$.} 
        \label{F5}
\end{figure} 

\subsection{Typical deviation of imbalance \label{aS1}}
%
In order to further get an estimate of the deviation of tdHF and non-interacting~(nI) charge imbalance from the exact imbalance $I(t)$, we define the following typical relative measure as,  
\eq{
\err{tdHF/nI} =  \exp\Le( \overline{\log\Le|\f{I_{\text{exact}}(t) - I_{\text{tdHF/nI}}(t)}{I_{\text{exact}}(t)}\Ri|}\Ri),
\label{eq:err}
}
where the overline implies averaging over disorder configurations. 
%
Figure~\ref{F5}(a,b) shows the dependence of $\err{tdHF}$ over $\err{nI}$ for weak interaction $U=0.25$ and for disorder strength $W=1.5, 5.0$ and Fig.~\ref{F5}(c,d) similar data for $U=1.0$. 

The objective behind showing the data in this way is to emphasize two points: a) we have seen previously~\cite{Bera2017} that in disordered interacting systems  time scales depend strongly on $W$;  therefore, it is often preferable to monitor the dynamics in terms of variables that emphasize the relevant timescales rather than $W$ itself. 
Our choice in Fig. \ref{F5} is to use the exact time evolution as a "clock" for occurances in non-interacting and tdHF dynamics. 
b) Such a way of representing the data further highlights the relative deviation of $\err{tdHF}$ in comparison to the non-interacting $\err{nI}$, which emphasizes effects of mean-field interactions. 

As one would expect, we observe that for small interaction, $U\lesssim \thop$,  and large disorder the typical deviation  $\err{tdHF}$ in absolute term is relatively small~($\sim 10-20\%$); this is further highlighted with the approach to the line of slope unity. At large disorder a vertical movement of deviation with increasing time is observedm, which reflects the approach towards an MBL phase ~(see Fig.~\ref{F5}(d)) and is consistent with the observation in Fig.~\ref{fig:avgImb}(b).

\section{Numerical details of RK4} \label{aTDHF}
\begin{figure}[b]
	\centering
	{\includegraphics[width=1.0\columnwidth]{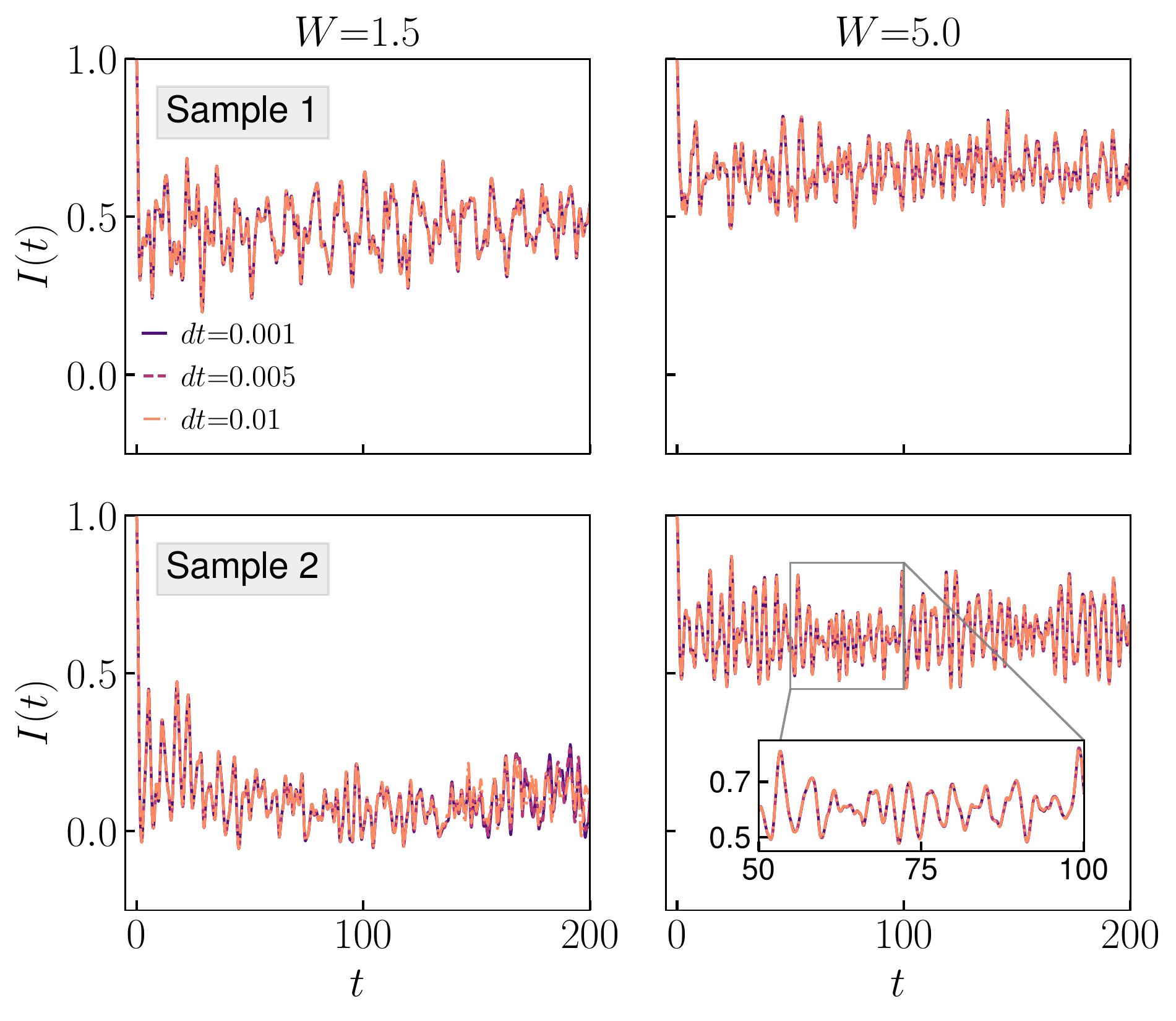}}
	\caption{Testing the convergence of RK4 time-integration with the time increment $dt$. Shown is $I(t)$ for two pairs of two samples, one with $W=1.5$ (left column) and one with $W=5.0$ (right column) at $U=1.0$. Three traces are shown per panel corresponding to $dt=0.01, 0.005, 0.001$ for RK4 integration for $L=16$. The inset highlights the fully converged behavior. } 
	\label{F11}
\end{figure}

In this section, we provide details of the numerical solution of the tdHF traces. To numerically solve Eq.~\eqref{Dyn2}, we resort to fourth order Runge-Kutta method (RK4) with uniform time step-size, $dt$. 
Figure~\ref{F11} shows a comparative study of the time steps ($dt=\{0.001, 0.005, 0.01\}$) for two individual sample as also shown in the main text. Within our simulation time the choice of $dt=0.01$ seems to be converged with respect to smaller $dt$. However, the timescale at which the RK4 integrator starts to acquire error depends on individual sample and also disorder strength. Such deviation is already seen in the second sample for $W=1.5$, see Fig.~\ref{F11}. At long time close to $t\sim 180 \thop^{-1}$ the convergence is poor. Therefore  we restrict our simulation time to $t_\text{max}=300$ to avoid such spurious convergence issues. 

\end{document}